\documentclass[10pt]{article}
\usepackage[margin=1in]{geometry}
\usepackage{graphicx}
\usepackage{wrapfig}
\usepackage{hyperref}
\hypersetup{
    colorlinks=true,
    linkcolor=blue,
    filecolor=magenta,
    urlcolor=cyan,
}
\usepackage{amsthm}
\usepackage{amsmath}
\usepackage{bm}
\usepackage{float}
\usepackage{nccmath}
\usepackage{longtable,tabularx}
\usepackage{caption}
\usepackage{dsfont}
\usepackage{ragged2e}
\graphicspath{{Pictures/}}
\usepackage{trimclip}
\def\AR{\clipbox{0pt 0pt .32em 0pt}\AE\kern-.30emR}

\title{\bfseries A Universal Optimal Control Strategy for a Tailsitter UAV}

\author{
    Animesh Kumar Shastry~\thanks{Postdoctoral Fellow, U.S. Army Research Laboratory. 
    This research was carried out entirely while the author was with the Department of Aerospace Engineering, IIT Kanpur.}
    \and
    Mangal Kothari~\thanks{Senior Principal  Flight Control Engineer, ADASI, EDGE Group, Abu Dhabi, UAE. 
    This research was carried out entirely while the author was with the Department of Aerospace Engineering, IIT Kanpur.}
}

\date{}

\begin{document}
	
\maketitle
	
\begin{abstract}
This work develops a unified optimal control framework for a Quadrotor Biplane tailsitter UAV capable of operating seamlessly across hover, transition, and cruise flight regimes. Although the tailsitter configuration enables mechanically simple mode switching, the transition maneuver remains challenging due to strong nonlinearities and rapidly varying aerodynamics. To address this, a trajectory optimization scheme based on nonlinear programming with direct collocation is formulated, incorporating nonlinear dynamics, actuator limits, and angle-of-attack constraints. The resulting optimal trajectories are safe, reliable, and time-efficient. For the cruise-to-hover maneuver, optimal trajectories are generated over a range of initial cruise velocities and subsequently learned using feedforward multilayer neural networks. The learned model generalizes across operating conditions and enables real-time generation of constraint-satisfying transition trajectories. These trajectories provide both feedforward control inputs and reference state profiles, which are tracked using a Model Predictive Controller (MPC). The MPC eliminates the need for controller switching or gain scheduling across flight envelopes, enabling a single universal controller for hover, transition, and cruise. A nonlinear Dynamic Inversion (DI) controller is also designed for comparison. Two numerical schemes for MPC are implemented and evaluated. Simulation results across all flight modes demonstrate that MPC achieves superior robustness to parameter uncertainties compared to DI. A computational cost analysis further highlights the trade-off between execution time and performance for the different MPC solvers.
\end{abstract}

\section{Introduction}
\label{sec:Introduction}

Unmanned Aerial Vehicles (UAVs) have evolved rapidly from simple aerial photography platforms into indispensable tools across industrial, civil, and defense sectors. Their ability to perform tasks such as surveying, mapping, cargo delivery, infrastructure inspection, and search-and-rescue has expanded expectations on range, endurance, and payload capability. However, conventional UAV configurations—fixed-wing, multi-rotor, and helicopter—each suffer from inherent limitations. Fixed-wing UAVs require runways or launch systems that are impractical in dense urban environments, while multi-rotor and helicopter platforms offer excellent hovering capability but poor cruise efficiency and limited endurance.

Hybrid Vertical Take-Off and Landing (VTOL) UAVs attempt to bridge this gap by combining the strengths of fixed-wing and rotary-wing aircraft. Among the various hybrid VTOL concepts—tilt-rotor, vector-thrust, and tail-sitter—the tail-sitter configuration stands out for its mechanical simplicity. A tail-sitter transitions between hover and forward flight by rotating its fuselage by approximately $90^\circ$, eliminating the need for additional actuators. Early tail-sitter designs such as the T-Wing~\cite{T-Wing} and the single-propeller R/C tail-sitter~\cite{simple_tailsitter} relied on aerodynamic control surfaces and prop-wash, resulting in limited hover performance but good stability in forward flight. To improve hovering capability without sacrificing cruise efficiency, researchers have explored quadrotor-based tail-sitter configurations.

A particularly compact and efficient variant is the Biplane Quadrotor tail-sitter, which uses two wings in a biplane arrangement to generate sufficient lift without the large wingspan required by single-wing designs. The foundational modeling and control of this configuration were established in a series of works by Kothari and co-authors~\cite{biplane_swiveling,biplane_transition,biplane_hinf,biplane_optimal,biplane_ILC}. These studies introduced the swiveling biplane-quadrotor concept~\cite{biplane_swiveling}, developed model-based transition control strategies~\cite{biplane_transition}, explored robust $H_\infty$ control~\cite{biplane_hinf}, and investigated optimal transition trajectories~\cite{biplane_optimal} as well as iterative learning-based feedforward control~\cite{biplane_ILC}. Collectively, this body of work demonstrated the aerodynamic advantages and maneuvering potential of the biplane configuration, while also highlighting the challenges posed by strong nonlinearities, post-stall behavior, and the absence of differential flatness.

These characteristics make traditional control approaches—often reliant on model invertibility or simplified aerodynamics—difficult to apply reliably. Recent advances in nonlinear control have improved performance for complex aerial vehicles, but accurate and differentiable aerodynamic models remain difficult to obtain, especially near stall~\cite{post-stall-modelling,post-stall-modelling2}. Many existing hybrid VTOL controllers rely on switching logic between hover and fixed-wing modes~\cite{VTOL-switching-control1,VTOL-switching-control2,VTOL-switching-control3} or gain-scheduled transition controllers~\cite{transition_controller,transition-gain-scheduling1,transition-gain-scheduling2}. These methods often depend on carefully tuned switching conditions and are not inherently optimal or robust. Tail-sitter-specific strategies remain limited, and reliable autonomous transition remains an open challenge.

Optimal control offers a principled way to generate dynamically feasible, constraint-satisfying transition trajectories. Direct methods discretize the problem and enforce dynamics and constraints at collocation points, while indirect methods rely on calculus of variations and lead to two-point boundary value problems. Although indirect methods can be elegant, they are often difficult to solve efficiently. Adaptive critic networks have been explored for real-time optimal control, but typically require large training datasets. For tail-sitters, the transition maneuver spans low-speed and high-speed aerodynamic regimes, making trajectory optimization particularly valuable.

In this work, we build upon the prior modeling and control foundations of the biplane quadrotor tailsitter~\cite{biplane_swiveling,biplane_transition,biplane_hinf,biplane_optimal,biplane_ILC} and focus on generating safe and reliable transition trajectories using a direct optimal control method with nonlinear programming and collocation. To enable real-time applicability, the optimized cruise-to-hover trajectories are further generalized using feedforward neural networks, allowing online generation of constraint-respecting trajectories for arbitrary initial cruise speeds.

To track these trajectories, we employ a Model Predictive Controller (MPC), which uses the system dynamics to predict future behavior and compute optimal control actions. MPC naturally handles constraints, provides disturbance rejection, and avoids the need for controller switching or gain scheduling across flight envelopes. Although MPC has been applied to various UAV platforms~\cite{Survey-MPC,Kim-MPC,MPC-Slegers,MPC-Bemporad,MPC-Kang}, its use in tail-sitters remains limited, particularly with emphasis on robustness to modeling uncertainties. For comparison, a nonlinear Dynamic Inversion (DI) controller with an outer–inner loop structure is also developed.

The main contributions of this work are:
\begin{itemize}
    \item development of a direct-method trajectory optimization framework for tail-sitter transition maneuvers with stall-aware constraints,
    \item generation and neural-network-based generalization of cruise-to-hover optimal trajectories for real-time use,
    \item design of a universal MPC controller capable of operating across hover, transition, and cruise without switching or gain scheduling,
    \item comparative evaluation of MPC and DI controllers under parameter uncertainties, and
    \item analysis of computational cost for different MPC numerical solvers.
\end{itemize}

The remainder of the paper is organized as follows. Section~\ref{sec:System Dynamics} presents the longitudinal dynamic model. Section~\ref{sec:Trajectory Optimization} describes the trajectory optimization framework and hover-to-cruise results. Section~\ref{sec:Trajectory Generalization} details the neural-network-based generalization of cruise-to-hover trajectories. Section~\ref{sec:Conclusion} concludes the paper and outlines directions for future work.

\section{System Dynamics}
\label{sec:System Dynamics}

\begin{figure}[h]
    \centering
    \begin{minipage}{0.5\textwidth}
        \centering
        \includegraphics[width=0.99\linewidth,clip,trim=0cm 0cm 5cm 0cm]{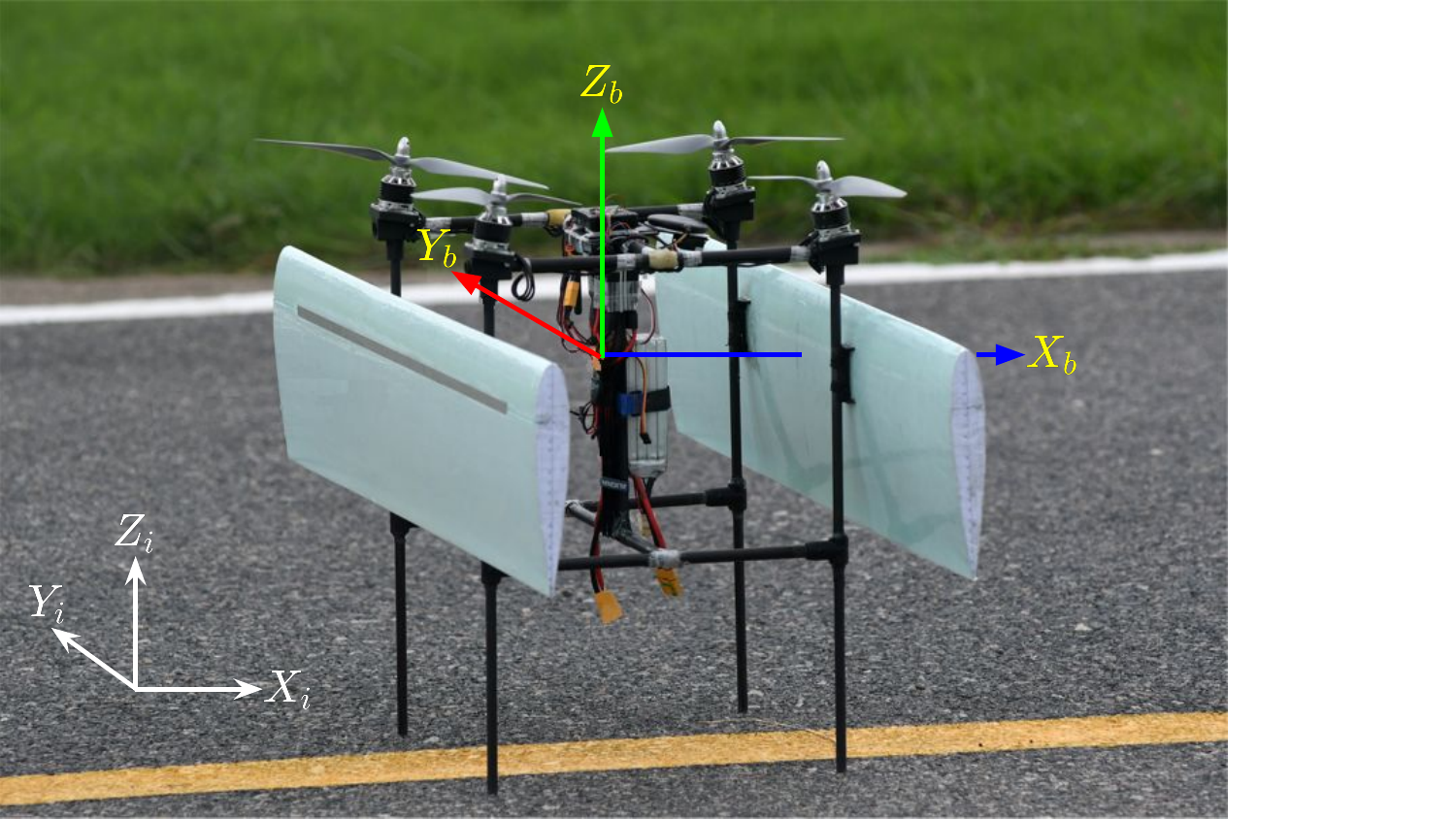}
        \captionsetup{width=.8\linewidth}
        \caption{Coordinate axes used for mathematical derivations.}
        \label{fig:coordinate_axes}
    \end{minipage}%
    \begin{minipage}{0.5\textwidth}
        \centering
        \includegraphics[width=0.99\linewidth,clip,trim=1.8cm 0cm 1.8cm 0cm]{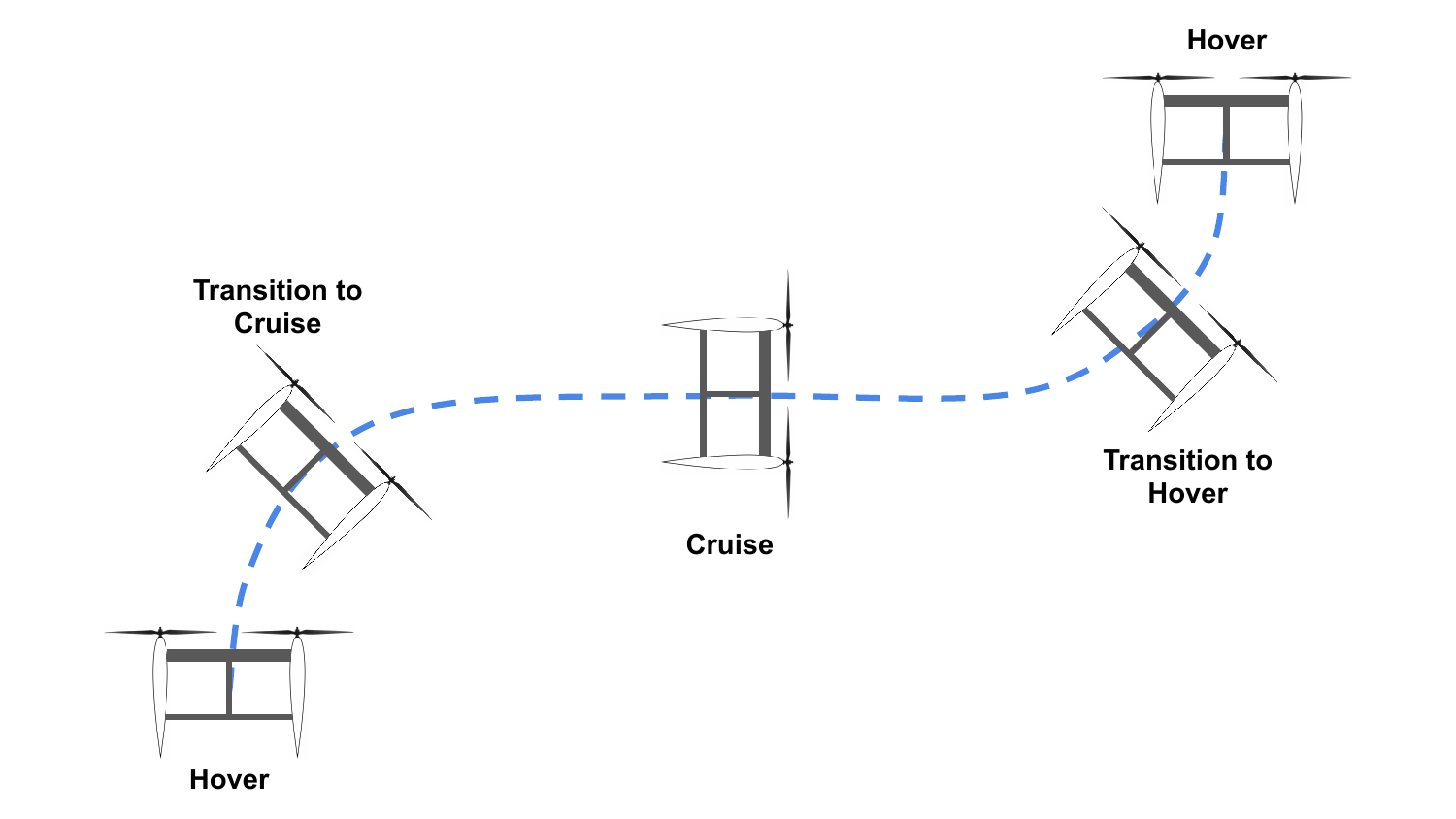}
        \captionsetup{width=.8\linewidth}
        \caption{Flight modes of a Biplane Quadrotor VTOL UAV.}
        \label{fig:Flight_Modes}
    \end{minipage}
\end{figure}

This section introduces the longitudinal dynamics of the Biplane Quadrotor Tailsitter UAV. Since the transition maneuver primarily occurs in the longitudinal plane, a reduced-order longitudinal model is used for trajectory generation and control design. A detailed 6-DOF aerodynamic and structural model of the vehicle can be found in~\cite{biplane_quadrotor_swati}. Figure~\ref{fig:coordinate_axes} illustrates the coordinate frames used for the derivation, and Figure~\ref{fig:Flight_Modes} shows the various flight modes experienced by the vehicle. Although the focus of this work is on transition flight, the controller developed later must remain valid across hover, transition, and cruise without switching or gain scheduling.

The vehicle is subjected to gravitational, aerodynamic, and propulsive forces, along with aerodynamic and propulsive moments. In the inertial $X_i$–$Z_i$ plane, the vehicle position is $[x\ z]^T$ and the velocity is $[\dot{x}\ \dot{z}]^T$. The pitch angle about the $Y_i$ axis is denoted by $\theta$, with pitch rate $\dot{\theta}$. The total thrust generated by the rotors is represented by the control input $u_1$, and the net pitching moment by $u_2$.

The compact form of the longitudinal dynamics is
\begin{equation}
    \dot{X} = f(X,U),
\end{equation}
where

\[
X = [x\ \dot{x}\ z\ \dot{z}\ \theta\ \dot{\theta}]^T, \qquad
U = [u_1\ u_2]^T.
\]

Expanding the dynamics yields
\begin{equation}
\label{dynamics_expanded}
f(X,U) =
\begin{bmatrix}
\dot{x} \\
\frac{1}{m}\left(u_1 \sin\theta + F_{ax}\cos\theta + F_{az}\sin\theta\right) \\
\dot{z} \\
\frac{1}{m}\left(u_1 \cos\theta - F_{ax}\sin\theta + F_{az}\cos\theta\right) - g \\
\dot{\theta} \\
\frac{1}{I_{yy}}\left(u_2 + \tau_a\right)
\end{bmatrix}.
\end{equation}

Here, $m$ is the vehicle mass, $g$ is gravitational acceleration, $I_{yy}$ is the pitch-axis moment of inertia, and $F_{ax}$, $F_{az}$, and $\tau_a$ are aerodynamic forces and moments. The aerodynamic forces in body axes are computed as
\begin{equation}
\label{AE_forces}
\begin{bmatrix}
F_{ax} \\
F_{az}
\end{bmatrix}
=
\begin{bmatrix}
\cos\alpha & \sin\alpha \\
-\sin\alpha & \cos\alpha
\end{bmatrix}
\begin{bmatrix}
-L \\
-D
\end{bmatrix},
\end{equation}
where $\alpha$ is the angle of attack, and $L$ and $D$ are the lift and drag forces:
\begin{equation}
\label{Lift_and_Drag}
\begin{bmatrix}
L \\
D
\end{bmatrix}
=
\frac{1}{2}\rho V^2 S
\begin{bmatrix}
C_L \\
C_D
\end{bmatrix}.
\end{equation}

Here, $\rho$ is air density, $V$ is the slipstream velocity, $S$ is the wing planform area, and $C_L$, $C_D$ are the lift and drag coefficients. Neglecting rotor downwash, the angle of attack and slipstream velocity simplify to

\[
\alpha = \frac{\pi}{2} - \theta - \tan^{-1}\left(\frac{\dot{z}}{\dot{x}}\right), \qquad
V = \sqrt{\dot{x}^2 + \dot{z}^2}.
\]

\begin{figure}[h]
    \centering
    \begin{minipage}{0.4\textwidth}
        \centering
        \includegraphics[width=0.99\linewidth]{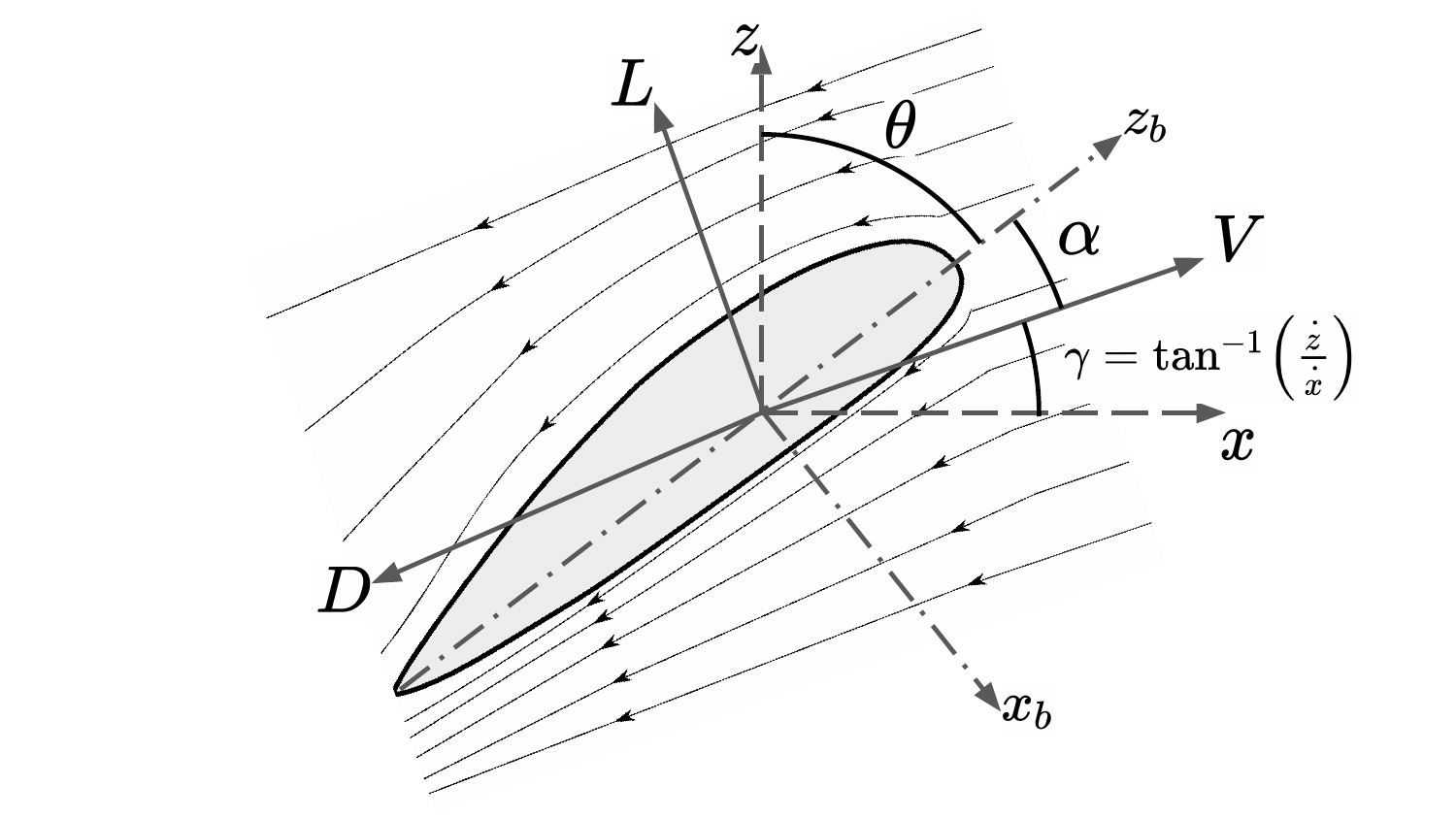}
        \captionsetup{width=.9\linewidth}
        \caption{Airfoil geometry and angle definitions.}
        \label{fig:airfoil}
    \end{minipage}%
    \begin{minipage}{0.6\textwidth}
        \centering
        \includegraphics[width=0.99\linewidth]{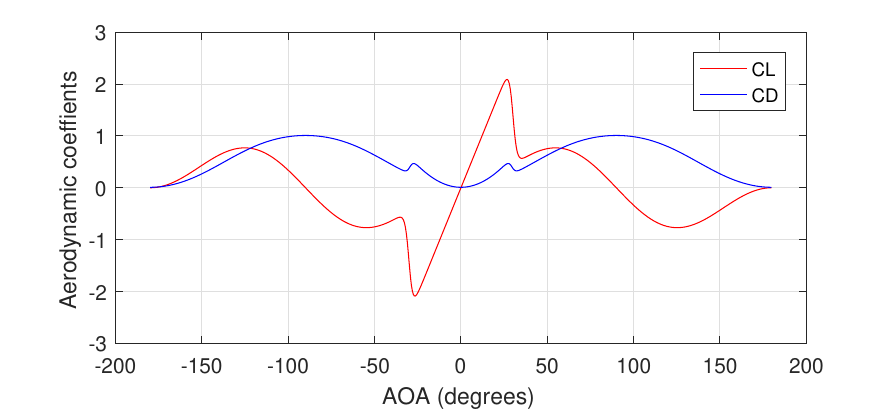}
        \captionsetup{width=.9\linewidth}
        \caption{Lift and drag coefficients vs.\ angle of attack.}
        \label{fig:CLandCD}
    \end{minipage}
\end{figure}

The aerodynamic coefficients are modeled as
\begin{equation}
\label{Lift_and_Drag_Coeff}
\begin{aligned}
C_D &= C_{D0} + k C_L^2 + \sin^2\alpha, \\
C_L &= C_L(\alpha) + C_{Lq}\frac{\dot{\theta}\bar{c}}{2V},
\end{aligned}
\end{equation}
where $\bar{c}$ is the mean aerodynamic chord. The pre- and post-stall lift behavior is captured using
\begin{equation}
\label{CLalpha}
C_L(\alpha) = (1-\sigma(\alpha))(C_{L0} + C_{L\alpha}\alpha)
+ \sigma(\alpha)\left[2\,\text{sign}(\alpha)\sin^2\alpha\cos\alpha\right],
\end{equation}
with the post-stall blending function
\begin{equation}
\label{sigma_alpha}
\sigma(\alpha) =
\frac{1 + e^{-M(\alpha-\alpha_{stall})} + e^{M(\alpha+\alpha_{stall})}}
     {\left(1 + e^{-M(\alpha-\alpha_{stall})}\right)\left(1 + e^{M(\alpha+\alpha_{stall})}\right)}.
\end{equation}

The aerodynamic pitching moment is modeled as
\begin{equation}
C_m = C_{m0} + C_{m\alpha}\alpha + C_{mq}\frac{\dot{\theta}\bar{c}}{2V},
\end{equation}
and the total aerodynamic moment acting on the vehicle is
\begin{equation}
\label{AE_moment}
\tau_a = \frac{1}{2}\rho V^2 S \bar{c} C_m + (z_{ac} - z_{cg})_b F_{ax}.
\end{equation}

\section{Trajectory Generation}
\label{sec:Trajectory Optimization}

This section describes the formulation and solution of the optimal control problem used to generate transition trajectories for the Biplane Quadrotor tailsitter. The goal is to compute an open-loop trajectory that minimizes a chosen cost function while satisfying system dynamics, actuator limits, and aerodynamic constraints. A direct collocation–based nonlinear programming (NLP) approach is adopted, which provides a flexible and numerically robust framework for handling nonlinear dynamics and constraints. Results for the hover-to-cruise transition are presented here, while the cruise-to-hover results and their generalization are discussed in Section~\ref{sec:Trajectory Generalization}.

\subsection{Direct Method}

\subsubsection{Nonlinear Programming}

The trajectory optimization problem is posed as a nonlinear program of the form
\begin{equation}
\begin{aligned}
\min_{Z} \quad & J(Z) \\
\text{subject to} \quad 
& c_{eq}(Z) = 0, \\
& c(Z) \le 0, \\
& Z_l \le Z \le Z_u,
\end{aligned}
\end{equation}
where $Z$ is the decision vector, $J(Z)$ is the cost function, $c_{eq}(Z)$ represents equality constraints (including system dynamics and boundary conditions), $c(Z)$ represents inequality constraints, and $Z_l$, $Z_u$ denote lower and upper bounds on the decision variables.

\subsubsection{Direct Collocation}

In the direct collocation method, the decision variables consist of the states and control inputs at a set of $N$ knot points, along with the final time $T_f$:
\begin{equation}
Z = [\, X_1\ X_2\ \ldots\ X_N\ U_1\ \ldots\ U_N\ T_f \,]^T,
\end{equation}
where $X_i \in \mathds{R}^6$ and $U_i \in \mathds{R}^2$ denote the state and control vectors at the $i^{\text{th}}$ knot.

The cost function is chosen as
\begin{equation}
J(Z) = Q_T T_f^2 + \sum_{i=0}^{N} \left( \bar{X}_i^T Q_X \bar{X}_i + U_i^T R U_i \right) h,
\end{equation}
where $h = T_f/(N-1)$ is the time step, $Q_T$ penalizes long transition times, $Q_X$ weights state deviations, and $R$ penalizes control effort. The deviation $\bar{X}_i$ is defined relative to the desired terminal state $X_{goal}$.

The weighting matrices are
\begin{equation}
Q_X = \begin{bmatrix}
0 & 0 & 0 & 0 & 0 & 0\\
0 & 1 & 0 & 0 & 0 & 0\\
0 & 0 & 1 & 0 & 0 & 0\\
0 & 0 & 0 & 1 & 0 & 0\\
0 & 0 & 0 & 0 & 50 & 0\\
0 & 0 & 0 & 0 & 0 & 50
\end{bmatrix}, \quad
R = \begin{bmatrix}
0.025 & 0 \\
0 & 0.11
\end{bmatrix}, \quad
Q_T = 9.
\end{equation}

\paragraph{Equality Constraints}

The equality constraints consist of the collocation dynamics and boundary conditions:
\begin{equation}
c_{eq}(Z) = [\, c_{dyn}(Z)\quad c_{bdr}(Z) \,]^T = 0.
\end{equation}

\begin{figure}[h]
    \centering
    \includegraphics[width=0.3\textwidth,clip,trim=0cm 0cm 1cm 0cm]{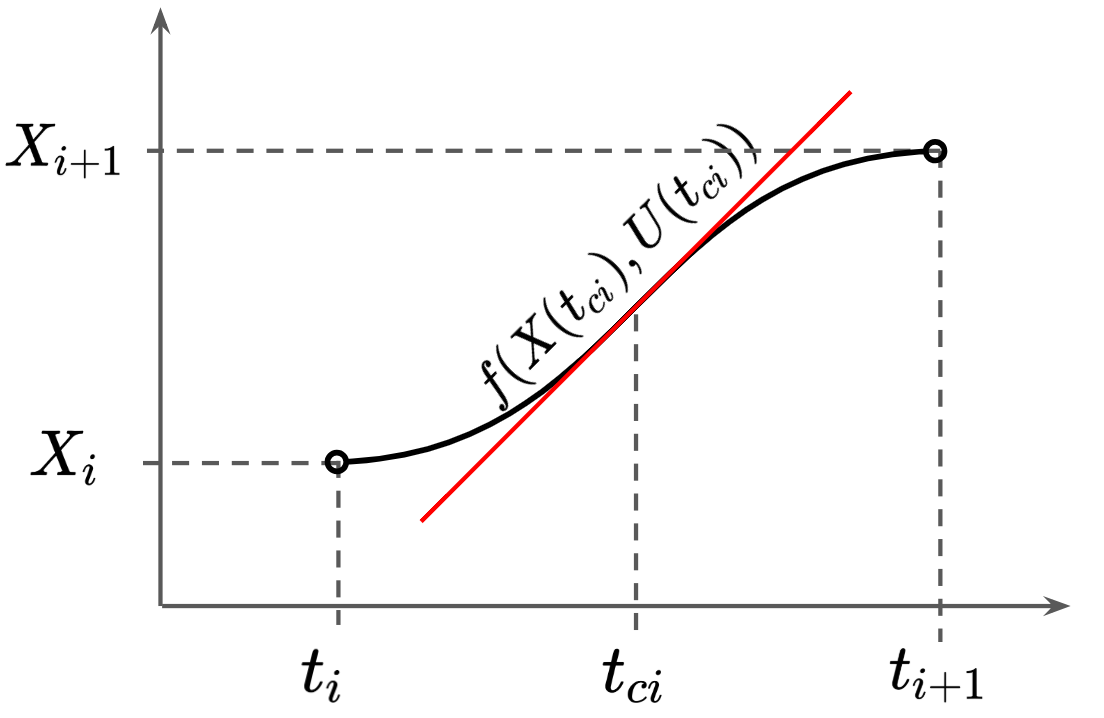}
    \caption{Collocation point between two adjacent knots.}
    \label{fig:collocation_pt}
\end{figure}

The collocation dynamics enforce consistency with the system model:
\begin{equation}
c_{dyn}(Z) = \dot{X}(t_{ci}) - f(X(t_{ci}), U(t_{ci})),
\end{equation}
where $t_{ci} = 0.5(t_i + t_{i+1})$ is the collocation point between two knots. These intermediate states are obtained using cubic interpolation. Figure~\ref{fig:collocation_pt} illustrates the concept.

Boundary constraints differ depending on the maneuver:

\paragraph{Cruise-to-hover:}
\begin{equation}
c_{bdr}(Z) =
\begin{bmatrix}
Z_{start} - Z_1 \\
Z_{goal} - [\dot{x}_N\ \dot{z}_N\ \theta_N\ \dot{\theta}_N\ U_N^T]^T
\end{bmatrix}.
\end{equation}

\paragraph{Hover-to-cruise:}
\begin{equation}
c_{bdr}(Z) =
\begin{bmatrix}
Z_{start} - Z_1 \\
Z_{goal} - [\dot{x}_N\ \dot{z}_N\ \dot{\theta}_N]^T
\end{bmatrix}.
\end{equation}

The position components of $Z_{goal}$ are omitted for cruise-to-hover transitions, since hover does not impose a fixed position. Similarly, position, attitude, and control components are removed for hover-to-cruise transitions.

\paragraph{Inequality Constraints}

To prevent stall, the angle of attack is constrained as
\begin{equation}
c(Z) = [c_0\ c_1\ \ldots\ c_N]^T, \qquad
c_i = |\alpha_i| - 0.6\,\alpha_{stall}.
\end{equation}

\paragraph{Bounds}

State and input bounds are imposed at all nodes:
\begin{equation}
\begin{aligned}
0 \le x_i \le 40, \qquad & 0 \le z_i \le 40, \\
-\pi \le \theta_i \le \pi, \qquad & 0 \le T_f, \\
0 \le u_{1,i} \le 0.8\,u_{1,max}, \qquad &
0.2\,u_{2,min} \le u_{2,i} \le 0.2\,u_{2,max}.
\end{aligned}
\end{equation}

The thrust and moment limits are intentionally scaled down to ensure sufficient margin for the feedback controller during trajectory tracking.

\paragraph{Numerical Solution}

\begin{figure}[h]
    \centering
    \includegraphics[width=0.7\linewidth]{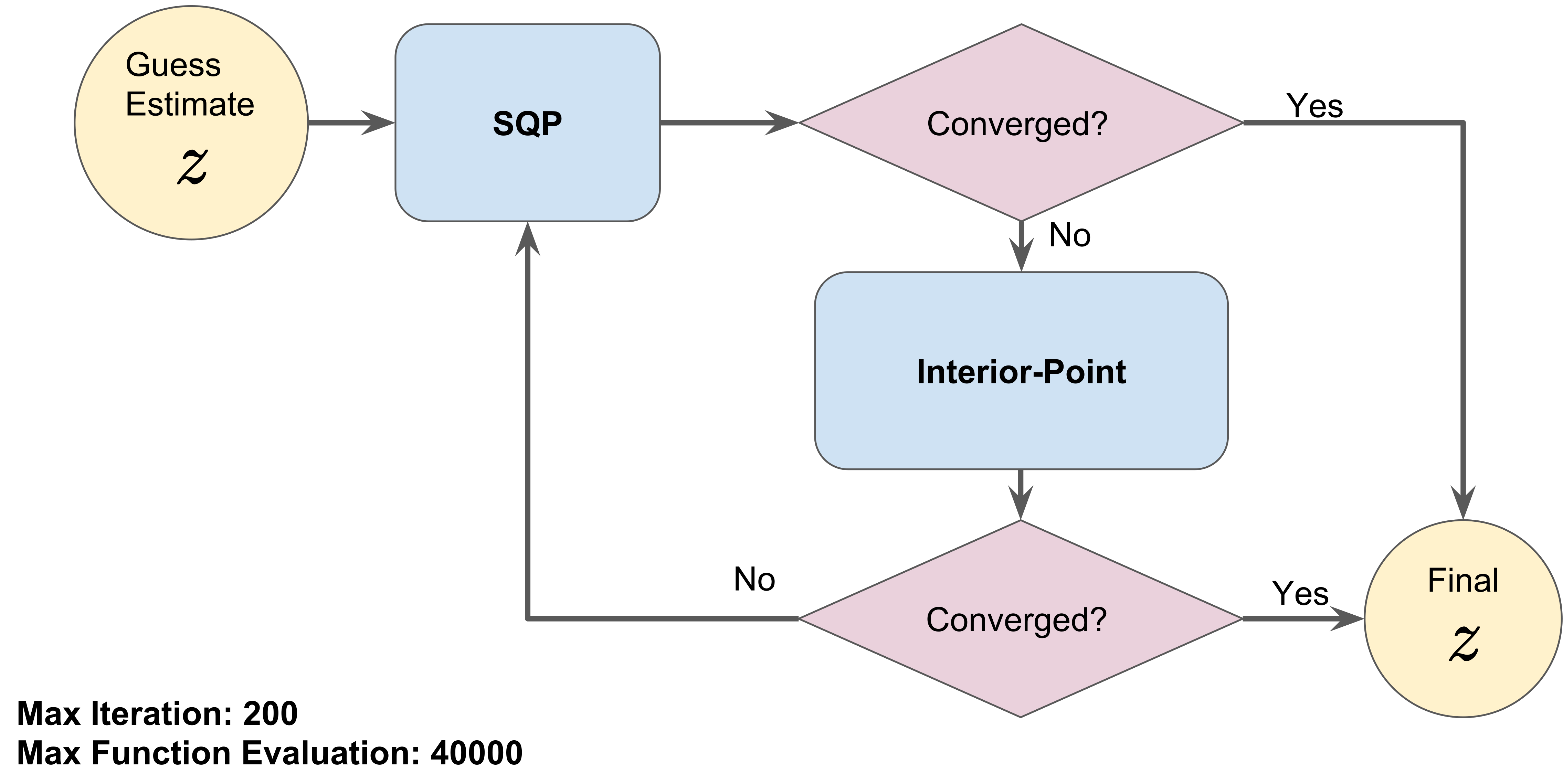}
    \caption{Nonlinear programming solution scheme.}
    \label{fig:NLP_Solve}
\end{figure}

The NLP is solved using \texttt{fmincon} (MATLAB), a gradient-based solver suitable for smooth nonlinear problems. As shown in Figure~\ref{fig:NLP_Solve}, two algorithms—Sequential Quadratic Programming (SQP) and Interior-Point—are used sequentially to improve convergence robustness.

\subsubsection{Hover-to-Cruise Transition Trajectory Optimization}

\begin{figure}[h]
    \centering
    \includegraphics[width=\linewidth,clip,trim=2.1cm 0cm 2.3cm 0cm]{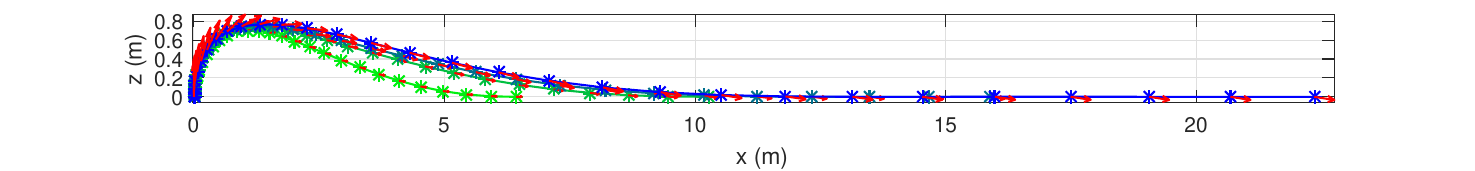}
    \caption{Planar view of optimized hover-to-cruise trajectories. Red arrows indicate thrust direction.}
    \label{fig:x_z_fwd}
\end{figure}

\begin{figure}[H]
    \centering
    \begin{minipage}{0.5\textwidth}
        \centering
        \includegraphics[width=\textwidth,clip,trim=0.5cm 1cm 0cm 1cm]{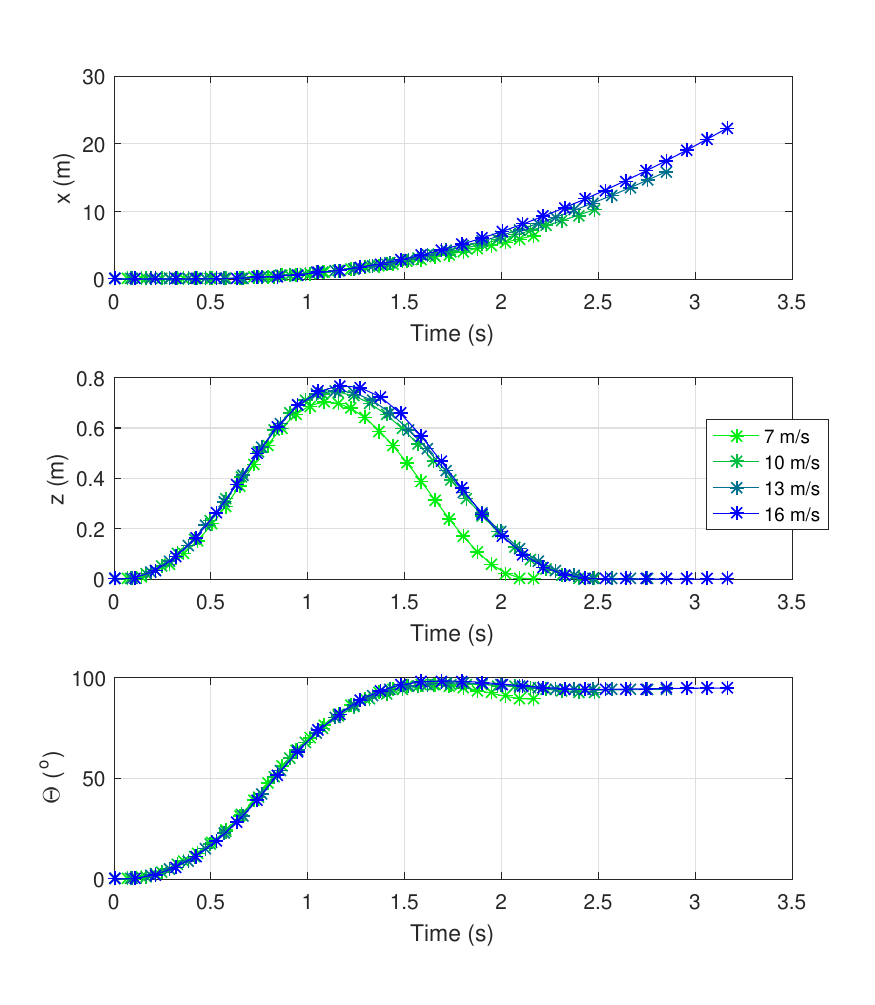}
    \end{minipage}%
    \begin{minipage}{0.5\textwidth}
        \centering
        \includegraphics[width=\textwidth,clip,trim=0.5cm 1cm 0cm 1cm]{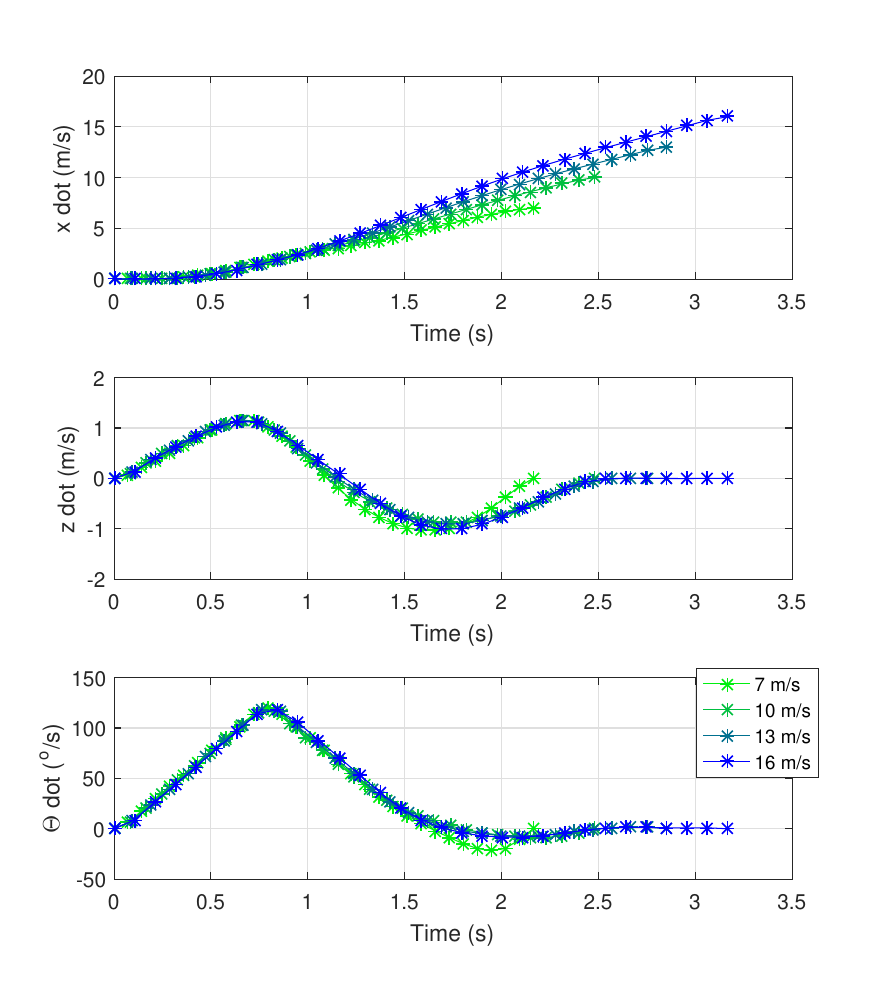}
    \end{minipage}
    \caption{State trajectories for different target cruise speeds.}
    \label{fig:state_fwd}
\end{figure}

\begin{figure}[H]
    \centering
    \begin{minipage}{0.45\textwidth}
        \centering
        \includegraphics[width=\linewidth,clip,trim=0.5cm 0cm 1cm 0.5cm]{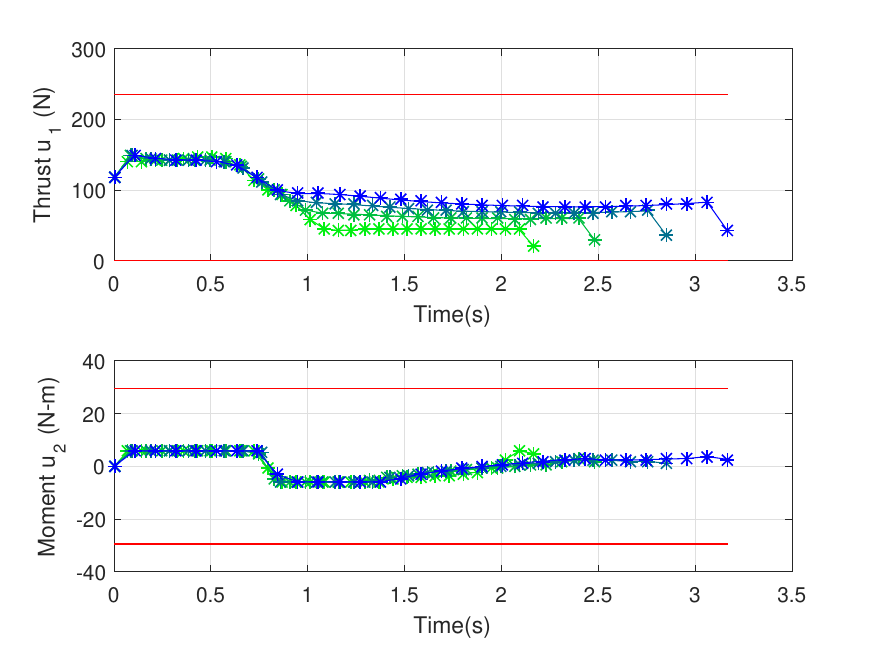}
        \caption{Control inputs along the optimized trajectories.}
        \label{fig:control_fwd}
    \end{minipage}%
    \begin{minipage}{0.55\textwidth}
        \centering
        \includegraphics[width=0.8\linewidth,clip,trim=0.5cm 0cm 1cm 0cm]{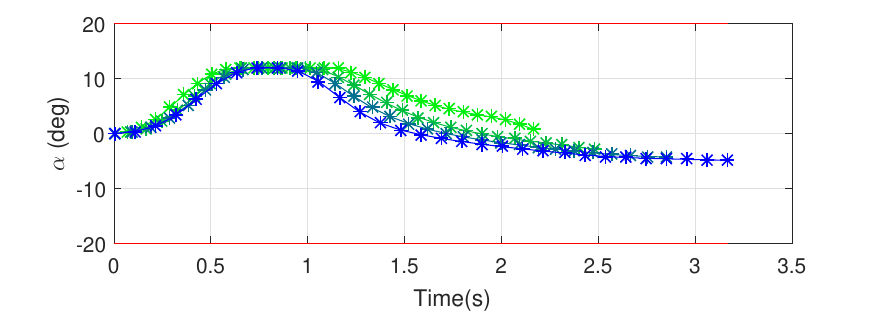}
        \caption{Angle of attack along the optimized trajectories.}
        \label{fig:AOA_fwd}
    \end{minipage}
\end{figure}

The direct collocation framework is used to compute hover-to-cruise transition trajectories for four target cruise speeds: 7, 10, 13, and 16 m/s. The optimization parameters are summarized in Table~\ref{tab:fwd_opt_par}. A planar view of the resulting trajectories is shown in Figure~\ref{fig:x_z_fwd}, followed by the state histories (Figure~\ref{fig:state_fwd}), control inputs (Figure~\ref{fig:control_fwd}), and angle-of-attack profiles (Figure~\ref{fig:AOA_fwd}). All trajectories satisfy the stall constraint of $20^\circ$.

\begin{table}[H]
    \centering
    \begin{tabular}{|c|c|}
        \hline
        \bf Parameter & \bf Value \\
        \hline\hline
        Number of knots ($N$) & 31 \\
        \hline
        $Z_{start}$ & $\begin{bmatrix} 0 & 0 & 0 & 0 & 0 & 0 & mg & 0 \end{bmatrix}^T$ \\
        \hline
        $Z_{goal}$ & $\begin{bmatrix} free & v_{cruise} & free & 0 & free & 0 & free & free \end{bmatrix}^T$ \\
        \hline
        $u_{1,max}$ & $2mg$ \\
        \hline
        $u_{2,max}$ & $0.25mg$ \\
        \hline
        $u_{2,min}$ & $-0.25mg$ \\
        \hline
    \end{tabular}
    \caption{Parameters for hover-to-cruise trajectory optimization.}
    \label{tab:fwd_opt_par}
\end{table}

\section{Trajectory Generalization}
\label{sec:Trajectory Generalization}

This section describes the generalization of optimal cruise-to-hover transition trajectories using a feedforward multilayer neural network. While the direct collocation method provides accurate optimal trajectories, solving the optimization problem online is computationally expensive. To enable real-time implementation on UAV hardware, a neural network is trained offline to approximate the mapping from initial cruise conditions to the corresponding optimal transition trajectory. The network takes selected boundary-condition parameters as inputs—primarily the initial cruise velocity—and outputs the full optimal trajectory.

\subsection{Feedforward Multi-layered ANN}

Feedforward neural networks are universal function approximators capable of representing any continuous mapping over a compact domain. This makes them well suited for learning the relationship between initial cruise conditions and the corresponding optimal transition trajectories. Although the hover-to-cruise maneuver is fully defined and does not benefit significantly from generalization, the cruise-to-hover maneuver depends strongly on the initial cruise speed. As a result, the neural network is trained specifically for the cruise-to-hover case, where the starting state varies and the optimal trajectory must adapt accordingly.

\subsection{Cruise-to-Hover Optimal Transition Trajectory Dataset Generation}

To construct the training dataset, the initial forward velocity $\dot{x}$ was varied uniformly from 5 to 20 m/s in increments of 1 m/s, yielding 16 distinct optimal trajectories. Each trajectory corresponds to a different initial cruise condition and is generated using the direct collocation method described earlier. The optimization parameters are summarized in Table~\ref{tab:bwd_opt_par}.

\begin{table}[H]
    \centering
    \begin{tabular}{|c|c|}
        \hline
        \bf Parameter & \bf Value \\
        \hline\hline
        Number of knots ($n$) & 31 \\
        \hline
        $Z_{start}$ & $\begin{bmatrix} 0 & v_{cruise} & 0 & 0 & \pi/2 & 0 & 0 & 0 \end{bmatrix}^T$ \\
        \hline
        $Z_{goal}$ & $\begin{bmatrix} free & 0 & free & 0 & 0 & 0 & mg & 0 \end{bmatrix}^T$ \\
        \hline
        $u_{1,max}$ & $2mg$ \\
        \hline
        $u_{2,max}$ & $0.25mg$ \\
        \hline
        $u_{2,min}$ & $-0.25mg$ \\
        \hline
    \end{tabular}
    \caption{Parameters for cruise-to-hover trajectory optimization.}
    \label{tab:bwd_opt_par}
\end{table}

\begin{figure}[H]
    \centering
    \begin{minipage}{0.5\textwidth}
        \centering
        \includegraphics[width=\textwidth,clip,trim=0.5cm 1cm 0cm 1cm]{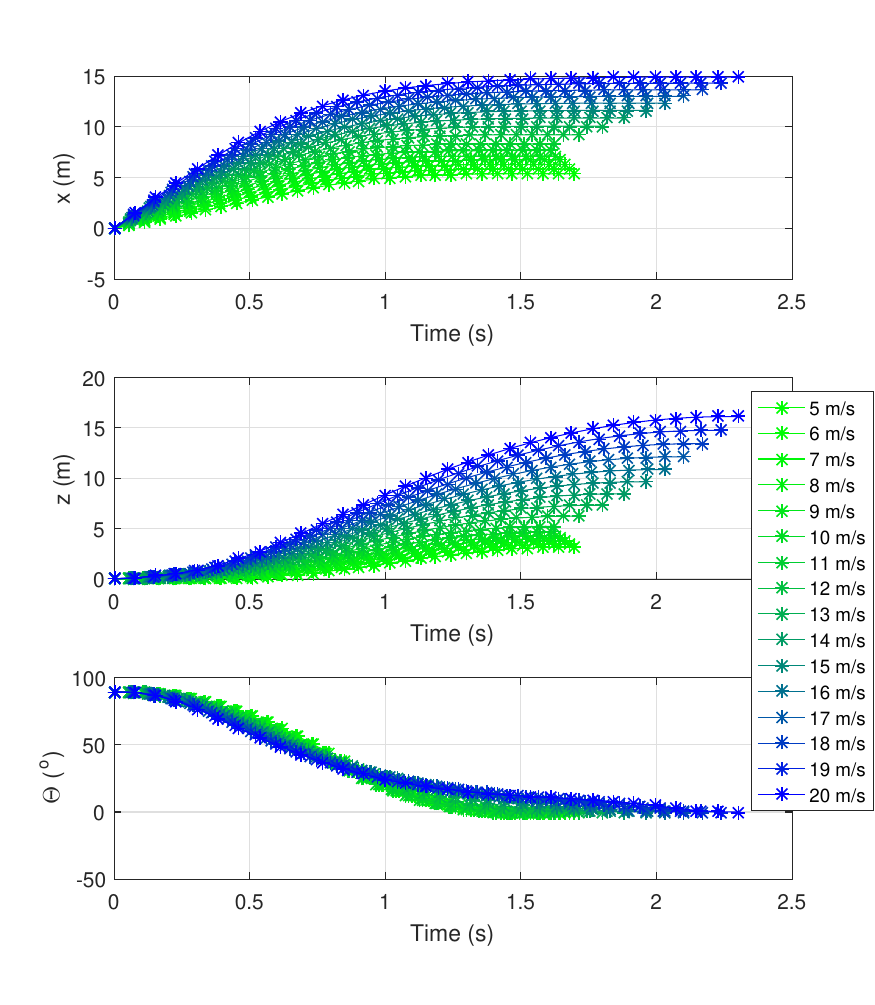}
    \end{minipage}%
    \begin{minipage}{0.5\textwidth}
        \centering
        \includegraphics[width=\textwidth,clip,trim=0.5cm 1cm 0cm 1cm]{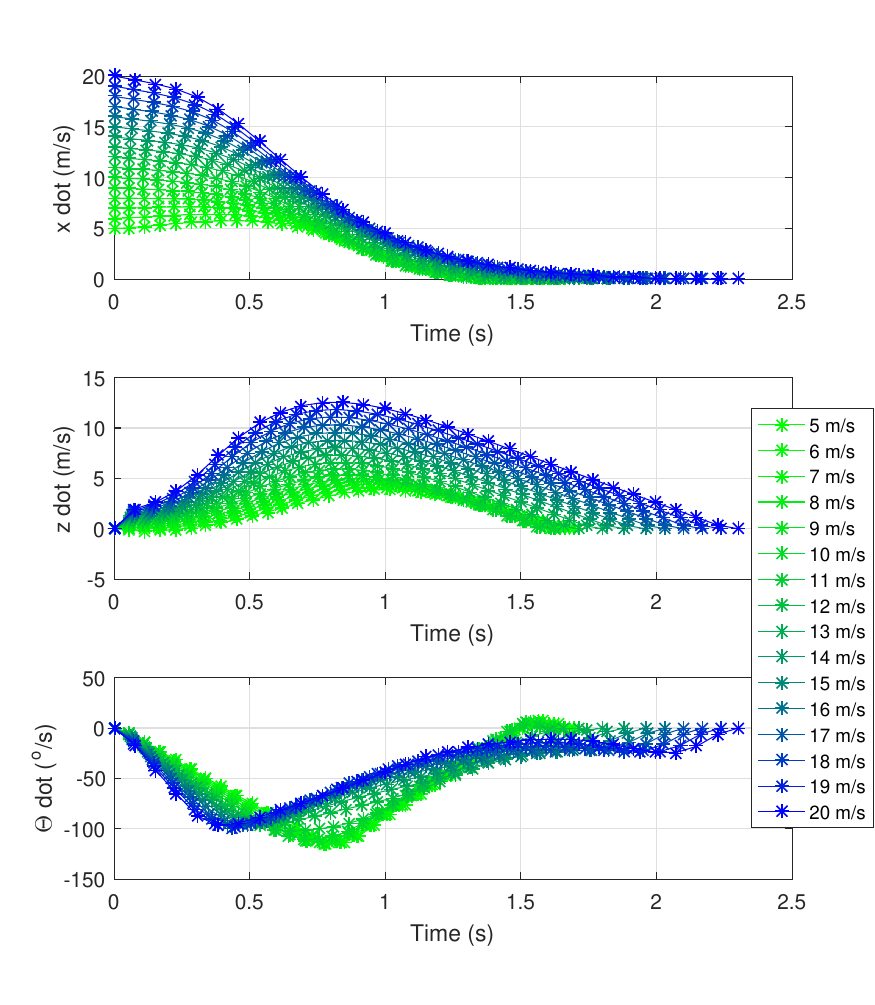}
    \end{minipage}
    \caption{State trajectories for the cruise-to-hover dataset. Different colors correspond to different initial cruise speeds.}
    \label{fig:state_dataset}
\end{figure}

\begin{figure}[H]
    \centering
    \begin{minipage}{0.45\textwidth}
        \centering
        \includegraphics[width=\linewidth]{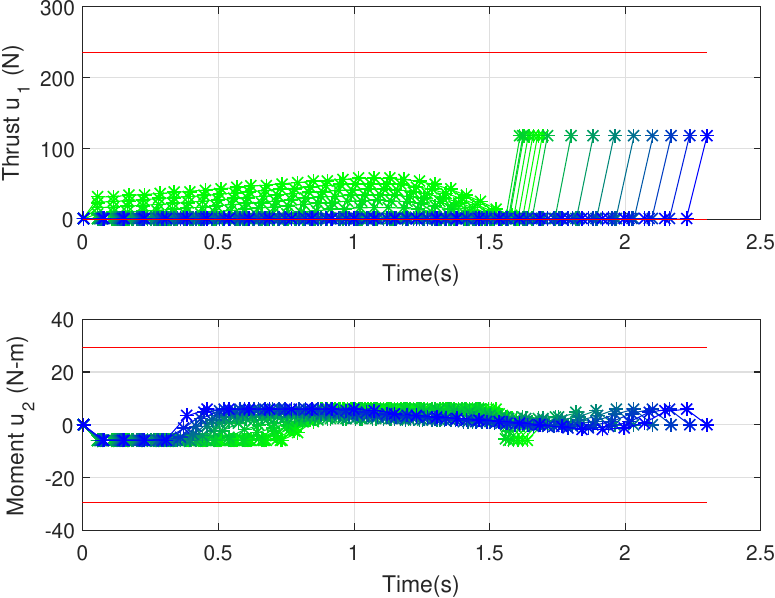}
        \caption{Control inputs for the trajectory dataset.}
        \label{fig:control_dataset}
        \includegraphics[width=\linewidth]{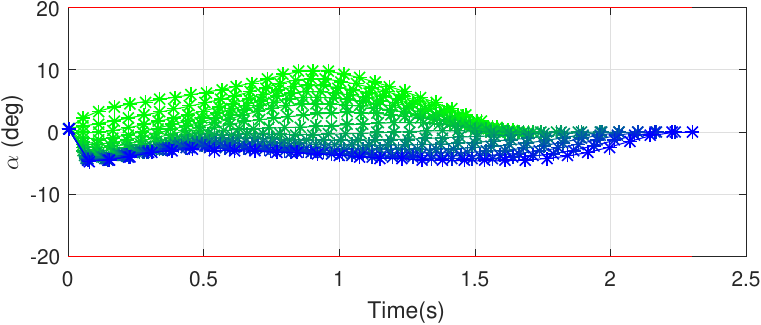}
        \caption{Angle of attack profiles for the trajectory dataset.}
        \label{fig:AOA_dataset}
    \end{minipage}%
    \begin{minipage}{0.55\textwidth}
        \centering
        \includegraphics[width=0.95\linewidth,clip,trim=0.8cm 1cm 1.2cm 1cm]{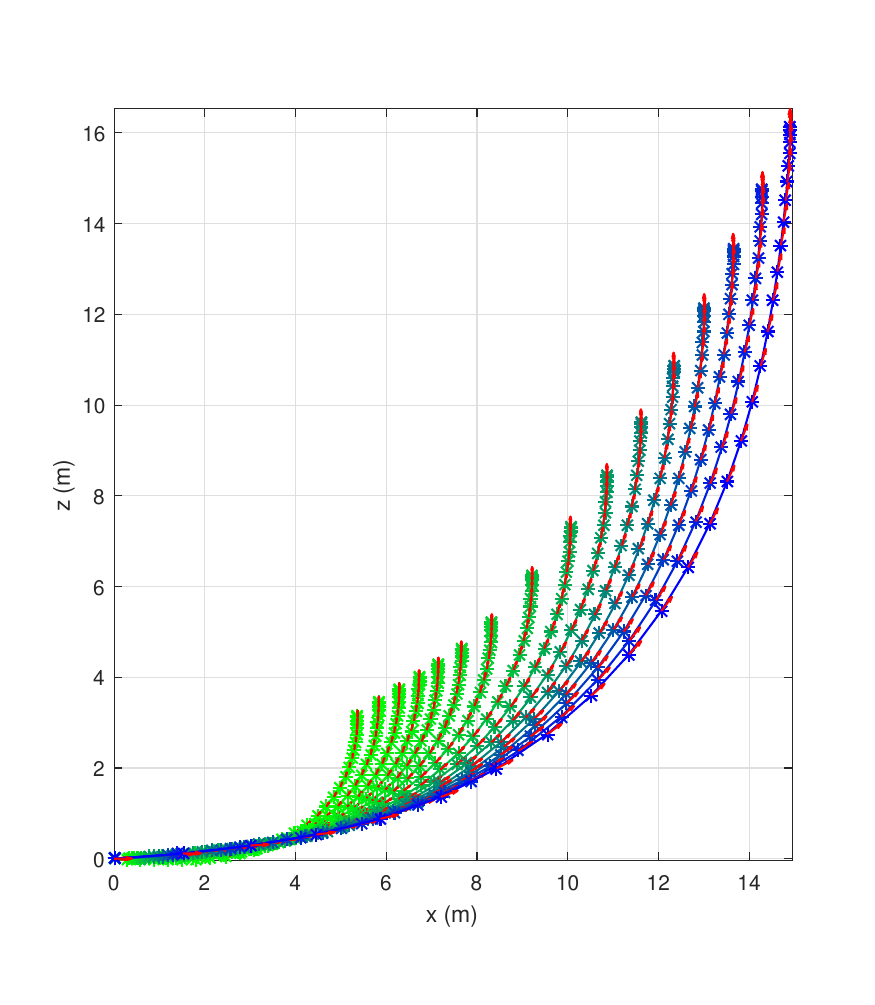}
        \captionsetup{width=.9\linewidth}
        \caption{Planar view of the cruise-to-hover trajectory dataset. Red arrows indicate thrust direction.}
        \label{fig:x_z_dataset}
    \end{minipage}
\end{figure}

The resulting dataset includes the full six-dimensional state trajectories and the corresponding control inputs, shown in Figures~\ref{fig:state_dataset} and~\ref{fig:control_dataset}. The angle of attack profiles (Figure~\ref{fig:AOA_dataset}) remain within the stall limit of $20^\circ$ for all trajectories. A planar view of the trajectories is provided in Figure~\ref{fig:x_z_dataset}, illustrating how the optimal transition path varies with initial cruise speed.

\subsection{Network Training}

\begin{figure}[H]
    \centering
    \begin{minipage}{0.5\textwidth}
        \centering
        \includegraphics[width=\linewidth]{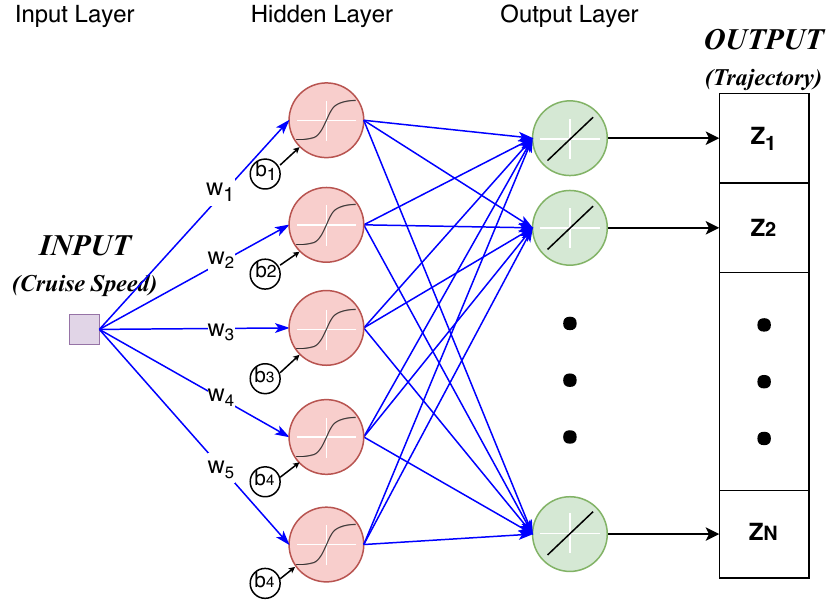}
        \captionsetup{width=.9\linewidth}
        \caption{Neural network architecture for generating transition trajectories from cruise speed.}
        \label{fig:ANN_Block}
    \end{minipage}%
    \begin{minipage}{0.5\textwidth}
        \centering
        \includegraphics[width=\linewidth]{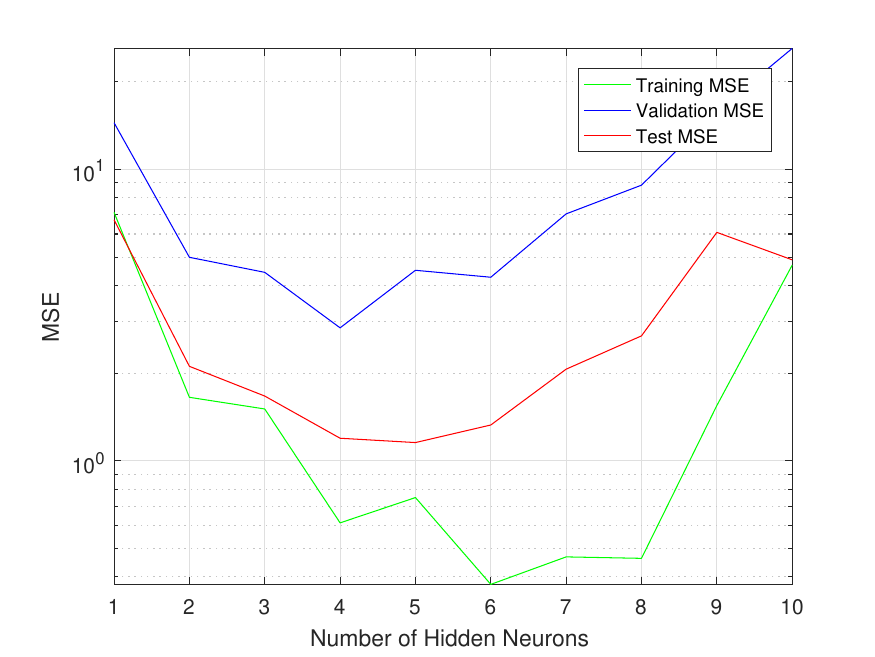}
        \captionsetup{width=.9\linewidth}
        \caption{Test MSE versus number of hidden neurons. Five neurons yield the best approximation accuracy.}
        \label{fig:MSE}
    \end{minipage}
\end{figure}

The neural network architecture used for trajectory generalization is shown in Figure~\ref{fig:ANN_Block}. The hidden layer employs hyperbolic tangent activation functions, while the output layer uses linear activations to represent the continuous trajectory values. Training is performed using the scaled conjugate gradient (SCG) algorithm~\cite{scg}, which avoids explicit line searches and is well suited for medium-sized regression problems.

To determine the appropriate network size, the number of hidden neurons is varied and the resulting test mean squared error (MSE) is evaluated. As shown in Figure~\ref{fig:MSE}, the minimum test MSE occurs with five hidden neurons, indicating that this architecture provides the best balance between accuracy and model complexity.

\section{Feedback Control Design}
\label{sec:Feedback Control Design}

To ensure that the UAV follows the optimal trajectories and rejects disturbances, a feedback control law is required. Such controllers use the desired state information supplied by the trajectory generator together with the current state estimates obtained from onboard sensors. The system dynamics are incorporated either explicitly, as in Dynamic Inversion (DI), or implicitly as a predictive model, as in Model Predictive Control (MPC). This section presents the development of both control strategies for the Biplane Quadrotor Tailsitter UAV and compares their performance through simulation.

\subsection{Multi-Loop Dynamic Inversion Control}

The Biplane Quadrotor Tailsitter is an underactuated system with coupled translational and rotational dynamics. A cascaded control structure is therefore adopted, where the outer loop regulates the slower translational motion and the inner loop stabilizes the faster rotational dynamics. The controller receives desired state variables ($\dot{x}_d$, $z_d$, $\dot{z}_d$, $\theta_d$, $\dot{\theta}_d$) and feedforward inputs ($T_{fwd}$, $m_{fwd}$) from the trajectory optimizer.

\subsubsection{Outer Loop}

The outer loop provides position-domain tracking by driving the errors in ($z - z_d$), ($\dot{x}_d - \dot{x}$), and ($\dot{z}_d - \dot{z}$) to zero, while regulating the lateral velocity $\dot{y}$ to zero. The $x$ and $y$ positions are not controlled directly due to sensor limitations (IMU and barometer only). The net acceleration command is computed as
\begin{equation}
acc_{net} = \frac{1}{m}R(0,\theta_d,0)\begin{bmatrix}0\\0\\T_{fwd}\end{bmatrix}
+ K_p \begin{bmatrix}0\\0\\z_d - z\end{bmatrix}
+ K_d \begin{bmatrix}\dot{x}_d - \dot{x}\\ -\dot{y}\\ \dot{z}_d - \dot{z}\end{bmatrix}
- \frac{1}{m}R_b^i F_a
+ \begin{bmatrix}0\\0\\g\end{bmatrix},
\end{equation}
where $K_p$ and $K_d$ are diagonal gain matrices, $F_a$ is the aerodynamic force in the body frame, and $R_b^i$ is the body-to-inertial rotation matrix.

The desired attitude is obtained from the commanded acceleration direction:
\begin{equation}
\begin{aligned}
b_3 &= \frac{acc_{net}}{\|acc_{net}\|}, \\
c_2 &= [-\sin(\psi_d)\ \cos(\psi_d)\ 0]^T, \\
b_1 &= \frac{c_2 \times b_3}{\|c_2 \times b_3\|}, \\
b_2 &= b_3 \times b_1, \\
R_d &= [\, b_1\ b_2\ b_3 \,].
\end{aligned}
\end{equation}

The thrust command is nominally
\begin{equation}
T = m\|acc_{net}\|.
\end{equation}
However, this always yields a positive thrust even when negative thrust would be ideal. A sign correction is introduced:
\begin{align}
flag &= \text{sign}\left([0\ 0\ 1]\, R_b^{iT} acc_{net}\right), \\
T &= flag \cdot m\|acc_{net}\|.
\end{align}
This correction is approximate and may chatter when attitude errors are large.

\subsubsection{Inner Loop}

The inner loop stabilizes the attitude using a geometric controller on $SO(3)$. The attitude kinematics and dynamics are
\begin{align}
\dot{R} &= R\hat{\omega}, \\
J\dot{\omega} &= -\omega \times J\omega + \tau + \tau_a,
\end{align}
where $J$ is the inertia matrix, $\tau_a$ is the aerodynamic torque, and $\tau$ is the rotor-generated control moment.

The configuration error function is
\begin{equation}
\Psi(R,R_d) = \frac{1}{2}\text{tr}(I - R_d^T R),
\end{equation}
with error vectors
\begin{align}
e_R &= \frac{1}{2}(R_d^T R - R^T R_d)^\vee, \\
e_\omega &= \omega - R^T R_d \omega_d.
\end{align}

The control moment is
\begin{equation}
\tau =
\begin{bmatrix}0\\ m_{fwd}\\ 0\end{bmatrix}
- K_R e_R - K_\omega e_\omega
+ \omega \times J\omega
- J(e_\omega \times R^T R_d \omega_d)
- \tau_a.
\end{equation}

\subsubsection{Block Diagram}

\begin{figure}[H]
\centering
\includegraphics[width=\linewidth]{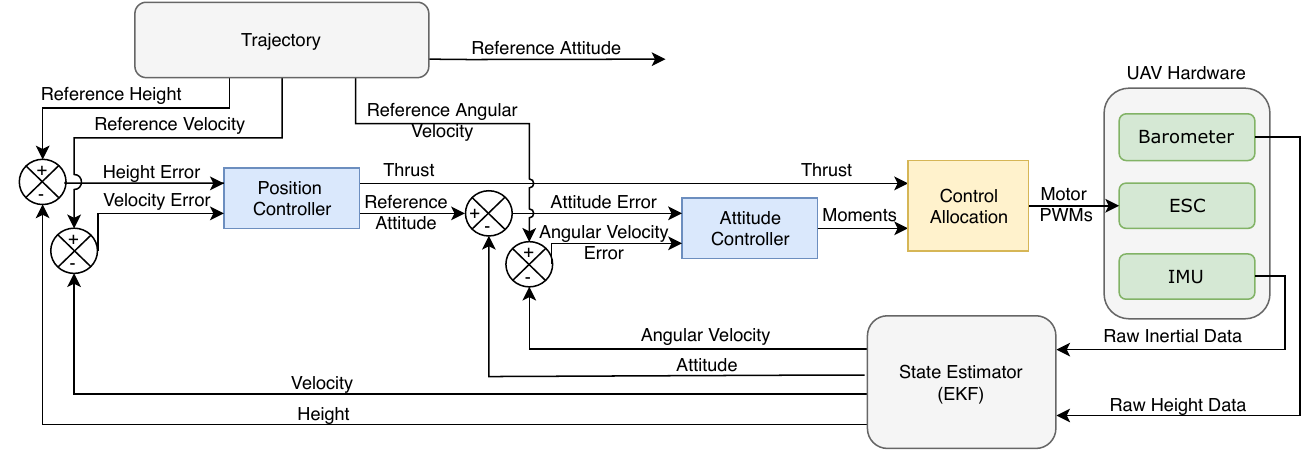}
\caption{Block diagram of the Dynamic Inversion feedback controller.}
\end{figure}

\subsubsection{Controller Issues}

\begin{itemize}
\item The reference trajectory is defined in the full state space, but DI uses only a subset (primarily position and attitude), leading to reduced tracking performance and robustness.
\item DI does not function reliably in forward flight for VTOL vehicles; cruise typically requires a separate controller, necessitating mid-flight switching.
\item The cascaded structure assumes fast attitude dynamics and slow translational dynamics, introducing delays and limiting achievable performance.
\end{itemize}

\subsection{Model Predictive Control}

MPC is an optimization-based control strategy that computes control actions by predicting future system behavior over a finite horizon. At each sampling instant $t$, an optimal control sequence is computed over a horizon $T = Nh$, but only the first control input is applied (receding horizon principle). The prediction model is the longitudinal dynamics described in Section~\ref{sec:System Dynamics}.

The predicted state and input sequences are
\begin{align}
X &= [X_0\ X_1\ \ldots\ X_N], \\
U &= [U_0\ U_1\ \ldots\ U_{N-1}],
\end{align}
with
\begin{align}
X_k &= X(t+kh), \\
U_k &= U(t+kh).
\end{align}

The discrete-time OCP is
\begin{equation}
(X^*,U^*) = \arg\min_{X,U} J(X,U)
\end{equation}
subject to
\begin{equation}
\begin{aligned}
X_0 &= X(t), \\
X_{k+1} &= X_k + \int_{t+kh}^{t+(k+1)h} f(X(\tau),U_k)\, d\tau, \\
\begin{bmatrix}0 \\ u_{2,min}\end{bmatrix}
&\le U_k \le
\begin{bmatrix}u_{1,max} \\ u_{2,max}\end{bmatrix}, \\
c_k(X_k) &< 0.
\end{aligned}
\end{equation}

The angle-of-attack constraint is enforced only in cruise:
\begin{equation}
c_k(X_k) =
\begin{cases}
|\alpha_k| - \alpha_{stall}, & V \ge 5\ \text{m/s}, \\
-1, & V < 5\ \text{m/s}.
\end{cases}
\end{equation}

The cost function is
\begin{equation}
J(X,U) = \frac{1}{2}\bar{X}_N^T Q_f \bar{X}_N
+ \frac{1}{2}\sum_{k=0}^{N-1}
\left(\bar{X}_k^T Q_X \bar{X}_k + U_k^T R U_k\right)h,
\end{equation}
where $\bar{X}_k = X_k - X_d(t+kh)$.

\subsubsection{MPC OCP Solving Schemes}

Two numerical schemes are implemented:

\paragraph{1) SQP-based Nonlinear Programming}

Sequential Quadratic Programming (SQP) approximates the Hessian of the Lagrangian using quasi-Newton updates (e.g., BFGS) and solves a sequence of QP subproblems. SQP handles constraints accurately but is computationally expensive due to repeated gradient and Hessian evaluations.

\paragraph{2) Indirect Method with Gradient Descent}

A simpler indirect method solves the OCP using calculus of variations and gradient descent. Control constraints are enforced via saturation after each update. This method is significantly faster but cannot enforce angle-of-attack constraints and therefore sacrifices some robustness.

The indirect OCP is
\begin{equation}
(X^*,U^*) = \arg\min_{X,U} \left[\Phi(X(t_f)) + \frac{1}{2}\int_{t_0}^{t_f} L(X,U)\, dt\right]
\end{equation}
subject to
\begin{equation}
X(t_0) = X_{start}, \qquad \dot{X}(t) = f(X(t),U(t)).
\end{equation}

The Hamiltonian is
\begin{equation}
H = L(X,U) + \lambda^T f(X,U),
\end{equation}
leading to the standard state, costate, and optimality conditions. The full expressions for the costate dynamics and the partial derivatives of the Hamiltonian 
with respect to the state and control variables are lengthy and are therefore provided in 
Appendix~\ref{appendix:derivatives}.

\section{Conclusion}
\label{sec:Conclusion}

Most nonlinear control strategies for aerial vehicles rely on dynamic inversion or related techniques that assume partial or full model invertibility. Hybrid VTOL platforms often depend on switching logic or gain scheduling to move between hover and forward flight, but such approaches are neither optimal nor inherently robust. In this work, we developed a unified optimal control strategy—based entirely on numerical methods—that enables a Biplane Quadrotor tailsitter UAV to operate seamlessly across hover, transition, and cruise flight envelopes. Trajectory optimization was performed for both cruise-to-hover and hover-to-cruise maneuvers using direct and indirect optimal control formulations. The optimization framework incorporated nonlinear aerodynamic effects, stall constraints on angle of attack, and actuator saturation limits. Because cruise-to-hover transitions may begin from a wide range of initial cruise speeds, a large dataset of optimal trajectories was generated and used to train a feedforward neural network. This enabled trajectory generalization and real-time generation of feasible transition trajectories for arbitrary cruise velocities. To track these trajectories, a Model Predictive Controller (MPC) was designed. MPC uses the predicted system response to compute optimal control inputs and naturally handles constraints, allowing a single controller to operate across all flight modes without switching or gain scheduling. For comparison, a nonlinear Dynamic Inversion (DI) controller with an inner–outer loop structure was also implemented. Two numerical schemes for MPC—one based on Sequential Quadratic Programming (SQP) and another based on Gradient Descent—were evaluated. Simulations across hover, transition, and cruise demonstrated that MPC consistently outperformed the DI controller. The DI controller performed adequately near hover but failed in cruise, whereas MPC maintained stable regulation and executed transitions reliably. MPC also showed superior robustness to parameter uncertainties, particularly during the transition maneuver. A computational cost comparison revealed that DI had the lowest execution time, followed by MPC with the Gradient Descent solver, and finally MPC with the SQP-based solver. However, when evaluated using the MPC cost function, the ranking reversed: the SQP-based MPC achieved the best performance, while the Gradient Descent variant offered the best balance between performance and computational efficiency. Although the Gradient Descent MPC could not enforce angle-of-attack constraints as effectively, it remains an attractive candidate for onboard implementation due to its low computational load.  Future work may include expanding the neural network inputs to incorporate additional state variables, enabling broader generalization of optimal transition trajectories. Practical implementation of the MPC controller on a real vehicle is another important direction, supported by appropriate onboard hardware capable of meeting the computational demands of MPC.

\bibliographystyle{unsrt}
\bibliography{Bibliography}

\appendix
\section{Derivatives for the Indirect MPC Solver}
\label{appendix:derivatives}

This appendix provides the full expressions required for the indirect optimal control method used in the MPC solver. These include the costate dynamics, the partial derivatives of the Hamiltonian with respect to the state and control variables, and the derivatives of the aerodynamic force and lift coefficient models.

\subsection{Hamiltonian and Necessary Conditions}

The Hamiltonian is defined as
\begin{equation}
H(X(t),U(t)) = L(X(t),U(t)) + \lambda(t)^T f(X(t),U(t)),
\end{equation}
where $L(X,U)$ is the instantaneous cost and $f(X,U)$ is the system dynamics. 
The necessary conditions for optimality are:
\begin{align}
\dot{X}(t) &= \frac{\partial H}{\partial \lambda(t)} = f(X(t),U(t)), \\
\dot{\lambda}(t) &= -\frac{\partial H}{\partial X(t)}, \\
\frac{\partial H}{\partial U(t)} &= 0, \\
X(t_0) &= X_{start}, \\
\lambda(t_f) &= \frac{\partial \Phi(X(t_f))}{\partial X(t_f)} = 2(X(t_f)-X_{goal})Q_X.
\end{align}

\subsection{Costate Dynamics}

The costate vector evolves according to
\begin{equation}
\dot{\lambda}(t) = -
\begin{bmatrix}
\frac{\partial H}{\partial x(t)} \\
\frac{\partial H}{\partial \dot{x}(t)} \\
\frac{\partial H}{\partial z(t)} \\
\frac{\partial H}{\partial \dot{z}(t)} \\
\frac{\partial H}{\partial \theta(t)} \\
\frac{\partial H}{\partial \dot{\theta}(t)}
\end{bmatrix}.
\end{equation}
The partial derivatives of the Hamiltonian are:
\paragraph{Derivative w.r.t. $x(t)$}
\begin{equation}
\frac{\partial H}{\partial x(t)} = 2(x(t)-x(t_f))Q_{X(1,1)}.
\end{equation}

\paragraph{Derivative w.r.t. $\dot{x}(t)$}
\begin{align*}
\frac{\partial H}{\partial \dot{x}(t)} &= 
2(\dot{x}(t)-\dot{x}(t_f))Q_{X(2,2)} + \lambda_1(t)
+ \frac{\lambda_6(t)}{I_{YY}}\frac{\partial \tau_a}{\partial \dot{x}(t)} \\
&\quad + \frac{1}{m}
\begin{bmatrix}
\lambda_2(t) & \lambda_4(t)
\end{bmatrix}
\begin{bmatrix}
\cos\theta(t) & \sin\theta(t) \\
-\sin\theta(t) & \cos\theta(t)
\end{bmatrix}
\begin{bmatrix}
\frac{\partial F_{ax}}{\partial \dot{x}(t)} \\
\frac{\partial F_{az}}{\partial \dot{x}(t)}
\end{bmatrix}.
\end{align*}

\paragraph{Derivative w.r.t. $z(t)$}
\begin{equation}
\frac{\partial H}{\partial z(t)} = 2(z(t)-z(t_f))Q_{X(3,3)}.
\end{equation}

\paragraph{Derivative w.r.t. $\dot{z}(t)$}
\begin{align*}
\frac{\partial H}{\partial \dot{z}(t)} &= 
2(\dot{z}(t)-\dot{z}(t_f))Q_{X(4,4)} + \lambda_3(t)
+ \frac{\lambda_6(t)}{I_{YY}}\frac{\partial \tau_a}{\partial \dot{z}(t)} \\
&\quad + \frac{1}{m}
\begin{bmatrix}
\lambda_2(t) & \lambda_4(t)
\end{bmatrix}
\begin{bmatrix}
\cos\theta(t) & \sin\theta(t) \\
-\sin\theta(t) & \cos\theta(t)
\end{bmatrix}
\begin{bmatrix}
\frac{\partial F_{ax}}{\partial \dot{z}(t)} \\
\frac{\partial F_{az}}{\partial \dot{z}(t)}
\end{bmatrix}.
\end{align*}

\paragraph{Derivative w.r.t. $\theta(t)$}
\begin{align*}
\frac{\partial H}{\partial \theta(t)} &= 
2(\theta(t)-\theta(t_f))Q_{X(5,5)} 
+ \frac{u_1(t)}{m}\left(\lambda_2(t)\cos\theta(t)-\lambda_4(t)\sin\theta(t)\right) \\
&\quad + \frac{1}{m}
\begin{bmatrix}
\lambda_2(t) & \lambda_4(t)
\end{bmatrix}
\begin{bmatrix}
\cos\theta(t) & \sin\theta(t) \\
-\sin\theta(t) & \cos\theta(t)
\end{bmatrix}
\begin{bmatrix}
\frac{\partial F_{ax}}{\partial \theta(t)} \\
\frac{\partial F_{az}}{\partial \theta(t)}
\end{bmatrix} \\
&\quad + \frac{1}{m}
\begin{bmatrix}
\lambda_2(t) & \lambda_4(t)
\end{bmatrix}
\begin{bmatrix}
-\sin\theta(t) & \cos\theta(t) \\
-\cos\theta(t) & -\sin\theta(t)
\end{bmatrix}
\begin{bmatrix}
F_{ax}(t) \\
F_{az}(t)
\end{bmatrix}.
\end{align*}

\paragraph{Derivative w.r.t. $\dot{\theta}(t)$}
\begin{align*}
\frac{\partial H}{\partial \dot{\theta}(t)} &= 
2(\dot{\theta}(t)-\dot{\theta}(t_f))Q_{X(6,6)} + \lambda_5(t)
+ \frac{\lambda_6(t)}{I_{YY}}\frac{\partial \tau_a}{\partial \dot{\theta}(t)} \\
&\quad + \frac{1}{m}
\begin{bmatrix}
\lambda_2(t) & \lambda_4(t)
\end{bmatrix}
\begin{bmatrix}
\cos\theta(t) & \sin\theta(t) \\
-\sin\theta(t) & \cos\theta(t)
\end{bmatrix}
\begin{bmatrix}
\frac{\partial F_{ax}}{\partial \dot{\theta}(t)} \\
\frac{\partial F_{az}}{\partial \dot{\theta}(t)}
\end{bmatrix}.
\end{align*}

\subsection{Aerodynamic Force Derivatives}

Using the aerodynamic force model
\[
\begin{bmatrix}
F_{ax} \\ F_{az}
\end{bmatrix}
=
\begin{bmatrix}
\cos\alpha & \sin\alpha \\
-\sin\alpha & \cos\alpha
\end{bmatrix}
\begin{bmatrix}
-L \\ -D
\end{bmatrix},
\]
The partial derivatives $\partial F_{ax}/\partial X$ and $\partial F_{az}/\partial X$ are:

\begin{align*}
\frac{\partial F_{ax}}{\partial \dot{x}} &= 
\frac{1}{2}\rho S \Big[
-2\dot{x}(C_L\cos\alpha - C_D\sin\alpha)
- \dot{z}(C_D\cos\alpha - C_L\sin\alpha) \\
&\quad - V^2(\cos\alpha + 2kC_L\sin\alpha)\frac{\partial C_L}{\partial \dot{x}}
- 2\sin^2\alpha\cos\alpha\,\dot{z}
\Big], \\
\frac{\partial F_{ax}}{\partial \dot{z}} &= 
\frac{1}{2}\rho S \Big[
-2\dot{z}(C_L\cos\alpha - C_D\sin\alpha)
+ \dot{x}(C_D\cos\alpha - C_L\sin\alpha) \\
&\quad - V^2(\cos\alpha + 2kC_L\sin\alpha)\frac{\partial C_L}{\partial \dot{z}}
+ 2\sin^2\alpha\cos\alpha\,\dot{x}
\Big], \\
\frac{\partial F_{ax}}{\partial \theta} &= 
\frac{1}{2}\rho V^2 S \Big[
- C_L\sin\alpha + C_D\cos\alpha + 2\sin^2\alpha\cos\alpha \\
&\quad - \frac{\partial C_L}{\partial \theta}(\cos\alpha + 2kC_L\sin\alpha)
\Big], \\
\frac{\partial F_{ax}}{\partial \dot{\theta}} &= 
-\frac{1}{4}\rho V S \bar{c} C_{Lq}(\cos\alpha + 2kC_L\sin\alpha),
\end{align*}

\begin{align*}
\frac{\partial F_{az}}{\partial \dot{x}} &= 
\frac{1}{2}\rho S \Big[
2\dot{x}(C_L\sin\alpha - C_D\cos\alpha)
+ \dot{z}(C_L\cos\alpha + C_D\sin\alpha) \\
&\quad + V^2(\sin\alpha - 2kC_L\cos\alpha)\frac{\partial C_L}{\partial \dot{x}}
- 2\sin\alpha\cos^2\alpha\,\dot{z}
\Big], \\
\frac{\partial F_{az}}{\partial \dot{z}} &= 
\frac{1}{2}\rho S \Big[
2\dot{z}(C_L\sin\alpha - C_D\cos\alpha)
- \dot{x}(C_L\cos\alpha + C_D\sin\alpha) \\
&\quad + V^2(\sin\alpha - 2kC_L\cos\alpha)\frac{\partial C_L}{\partial \dot{z}}
+ 2\sin\alpha\cos^2\alpha\,\dot{x}
\Big], \\
\frac{\partial F_{az}}{\partial \theta} &= 
\frac{1}{2}\rho V^2 S \Big[
- C_L\cos\alpha - C_D\sin\alpha + 2\sin\alpha\cos^2\alpha \\
&\quad + \frac{\partial C_L}{\partial \theta}(\sin\alpha - 2kC_L\cos\alpha)
\Big], \\
\frac{\partial F_{az}}{\partial \dot{\theta}} &= 
\frac{1}{4}\rho V S \bar{c} C_{Lq}(\sin\alpha - 2kC_L\cos\alpha).
\end{align*}

\subsection{Lift Coefficient Derivatives}

Using the lift model
\[
C_L = C_L(\alpha) + C_{Lq}\frac{\dot{\theta}\bar{c}}{2V},
\]
The derivatives are:
\begin{align*}
\frac{\partial C_L}{\partial \dot{x}} &= 
- C_{Lq}\frac{\dot{\theta}\bar{c}\dot{x}}{2V^3}
+ (1-\sigma)C_{L\alpha}\frac{\dot{z}}{V^2}
+ \frac{\partial\sigma}{\partial \dot{x}}(\cdots)
+ \sigma\frac{\dot{z}}{V^2}(\cdots), \\
\frac{\partial C_L}{\partial \dot{z}} &= 
- C_{Lq}\frac{\dot{\theta}\bar{c}\dot{z}}{2V^3}
- (1-\sigma)C_{L\alpha}\frac{\dot{x}}{V^2}
+ \frac{\partial\sigma}{\partial \dot{z}}(\cdots)
- \sigma\frac{\dot{x}}{V^2}(\cdots), \\
\frac{\partial C_L}{\partial \theta} &= 
-(1-\sigma)C_{L\alpha}
+ \frac{\partial\sigma}{\partial \theta}(\cdots)
- \sigma(\cdots).
\end{align*}

\subsection{Post-Stall Compensator Derivatives}

The compensator
\[
\sigma(\alpha) = 
\frac{1 + e^{-M(\alpha-\alpha_{stall})} + e^{M(\alpha+\alpha_{stall})}}
{(1+e^{-M(\alpha-\alpha_{stall})})(1+e^{M(\alpha+\alpha_{stall})})}
\]
has derivatives:

\begin{align*}
\frac{\partial \sigma}{\partial \dot{x}} &= 
\frac{M\dot{z}}{V^2}\frac{(\cdots)}{(1+e^{-M(\alpha-\alpha_{stall})})^2(1+e^{M(\alpha+\alpha_{stall})})^2}, \\
\frac{\partial \sigma}{\partial \dot{z}} &= 
-\frac{M\dot{x}}{V^2}\frac{(\cdots)}{(1+e^{-M(\alpha-\alpha_{stall})})^2(1+e^{M(\alpha+\alpha_{stall})})^2}, \\
\frac{\partial \sigma}{\partial \theta} &= 
-M\frac{(\cdots)}{(1+e^{-M(\alpha-\alpha_{stall})})^2(1+e^{M(\alpha+\alpha_{stall})})^2}.
\end{align*}

\end{document}